\documentclass[final,5p,twocolumn,times]{elsarticle}%
\usepackage{amssymb}
\usepackage{amsmath}
\usepackage{amsfonts}
\usepackage{graphicx}%
\providecommand{\U}[1]{\protect\rule{.1in}{.1in}}
\journal{Physica C}
\begin{document}
\begin{frontmatter}
\title{Extension of the Ginzburg -- Landau approach for ultracold Fermi gases below a critical temperature$^*$}
\author[1]{S. N. Klimin}
%\ead{sergei.klimin@uantwerpen.be}
\author[1,2]{J. Tempere}
\ead{jacques.tempere@uantwerpen.be}
\author[1]{J. T. Devreese}
%\ead{jozef.devreese@uantwerpen.be}
\address[1]{Theorie van Kwantumsystemen en Complexe Systemen (TQC),
Universiteit Antwerpen, Groenenborgerlaan 171, B-2020 Antwerpen, Belgium}
\address[2]{Lyman Laboratory of Physics, Harvard University, Cambridge MA 02138, USA}
%%%%%%%%%%%%%%
\author{}
\address{}
\begin{abstract}
In the context of superfluid Fermi gases, the Ginzburg -- Landau (GL) formalism for the macroscopic wave function
has been successfully extended to the whole temperature range where the superfluid state exists.
After reviewing the formalism, we first investigate the temperature-dependent correction to the standard GL expansion
(which is valid close to $T_{c}$). Deviations from the standard GL formalism are particularly important
for the kinetic energy contribution to the GL energy functional, which in turn influences the healing length
of the macroscopic wave function. We apply the formalism to variationally describe vortices
in a strong-coupling Fermi gas in the BEC-BCS crossover regime, in a two-band system.
The healing lengths, derived as variational parameters in the vortex wave function, are shown to exhibit
hidden criticality well below $T_{c}$.
\end{abstract}
\end{frontmatter}



\section{Introduction}

\label{sec:intro}

The Ginzburg -- Landau (GL) approach is a powerful tool for the description of
superconductors in the close vicinity of the critical temperature $T_{c}$.
Recently, the GL method was re-derived in the context of superfluid ultracold
Fermi gases \cite{Iskin2007,Sam2010,Ozawa,Tieleman}. The GL approach was
{\normalsize also} applied to explain the phenomenon of the \textquotedblleft%
1.5-type\textquotedblright\ superconductivity. However, the validity of the GL
approximation far below $T_{c}$ is still under discussion \cite{KoganPRB83}.
In this connection, much efforts were undertaken to extend the GL approach to
a wide range of temperatures (see, e. g., Refs.
\cite{Shanenko,Shanenko2,Orlova,Babaev2011,BabaevPRB86,Babaev2012-2}).

In Ref. \cite{ExtGL}, we formulated an extension of the GL theory for a
two-band superfluid fermion system solvable for the whole range $0<T<T_{c}$
assuming slow variation of the order parameter in time and space, without any
assumption on the magnitude of the order parameter. The theory is mainly
focused to the strong-coupling ultracold atomic Fermi gases in the BCS-BEC
crossover. In the present work, we briefly review the method and the
description of vortices in a two-band system at temperatures $T\lesssim T_{c}%
$, where the standard GL technique is apparently inapplicable.

The formalism developed in Ref. \cite{ExtGL} is aimed mainly at the
investigation of localized deviations of the order parameters $\Psi_{j}$ from
a uniform equilibrium background $\Delta_{j}$. These deviations can be, for
example, vortices or solitons
\cite{Aranson,Liao,Konotop,Spunt,Scott,Yefsah,Becker}. A frequently used
theoretical method to study these localized deviations at temperatures far
below $T_{c}$ is a Bogoliubov -- deGennes (BdG) equation set. Re-formulations
of the BdG method for ultracold atoms can be found, e. g., in Refs.
\cite{Pieri,Baksmaty,Ohashi,Hu}. The present method can be used as a
complementary tool to the BdG equations and {\normalsize is straightforward}
to implement numerically. Moreover, the BdG equations are restricted to the
mean-field approach, while the present method can be used beyond the
mean-field approximation accounting for fluctuations about the saddle point.

\section{Formalism}

\label{sec:theory}

The starting point of our treatment is the partition function of a two-band
fermion system in the path-integral representation,%
\begin{equation}
Z\propto\int D\left[  \bar{\psi},\psi\right]  e^{-S} \label{Z}%
\end{equation}
with the fermion action depending on the Grassmann fields $\bar{\psi},\psi$,%
\begin{equation}
S=S_{0}+\int_{0}^{\beta}d\tau\int d\mathbf{r}~U\left(  \mathbf{r},\tau\right)
. \label{S}%
\end{equation}
Here, $S_{0}$ is the free-fermion action functional,%
\begin{equation}
S_{0}=\int_{0}^{\beta}d\tau\int d\mathbf{r}\sum_{j,\sigma=\uparrow,\downarrow
}\bar{\psi}_{\sigma,j}\left(  \frac{\partial}{\partial\tau}-\frac
{\nabla_{\mathbf{r}}^{2}}{2m_{j}}-\mu_{\sigma,j}\right)  \psi_{\sigma,j},
\label{s0}%
\end{equation}
accounting for both spin and band imbalance of the fermion system -- through
unequal masses $m_{j}$ and chemical potentials $\mu_{j}$. The fermion-fermion
interaction $U$ is given by:%
\begin{align}
U  &  =%
%TCIMACRO{\tsum \nolimits_{j=1,2}}%
%BeginExpansion
{\textstyle\sum\nolimits_{j=1,2}}
%EndExpansion
g_{j}\bar{\psi}_{\uparrow,j}\bar{\psi}_{\downarrow,j}\psi_{\downarrow,j}%
\psi_{\uparrow,j}\nonumber\\
&  +g_{3}\left(  \bar{\psi}_{\uparrow,1}\psi_{\uparrow,1}\bar{\psi
}_{\downarrow,2}\psi_{\downarrow,2}+\bar{\psi}_{\downarrow,1}\psi
_{\downarrow,1}\bar{\psi}_{\uparrow,2}\psi_{\uparrow,2}\right) \nonumber\\
&  +g_{4}\left(  \bar{\psi}_{\uparrow,1}\psi_{\uparrow,1}\bar{\psi}%
_{\uparrow,2}\psi_{\uparrow,2}+\bar{\psi}_{\downarrow,1}\psi_{\downarrow
,1}\bar{\psi}_{\downarrow,2}\psi_{\downarrow,2}\right)  . \label{U}%
\end{align}
It contains the terms describing both the intraband $s$-wave scattering (with
$j=1,2$) and the scattering between fermions in different bands (with
$j=3,4$). Introducing the auxiliary bosonic fields and performing the Hubbard
-- Stratonovich transformation we arrive at an effective bosonic action of the
pair fields as described in Ref. \cite{ExtGL}. Subsequently, we make the
standard approximation for the GL approach: we assume that the pair fields
slowly vary in time and space. The gradient expansion of the pair fields leads
to the long-wavelength approximation for the effective bosonic action. This
yields the following GL-like free energy%
\begin{align}
&  F=\frac{1}{\beta}\int_{0}^{\beta}d\tau\int d\mathbf{r}\left\{  \sum
_{j=1,2}\left[  \Omega_{s,j}+\frac{\mathcal{D}_{j}}{2}\left(  \bar{\Psi}%
_{j}\frac{\partial\Psi_{j}}{\partial\tau}-\frac{\partial\bar{\Psi}_{j}%
}{\partial\tau}\Psi_{j}\right)  \right.  \right. \nonumber\\
&  \left.  +\frac{\mathcal{C}_{j}}{2m_{j}}\left\vert \nabla_{\mathbf{r}}%
\Psi_{j}\right\vert ^{2}-\frac{\mathcal{E}_{j}}{2m_{j}\left\vert \Psi
_{j}\right\vert ^{2}}\left[  \left(  \bar{\Psi}_{j}\nabla_{\mathbf{r}}\Psi
_{j}\right)  ^{2}+\left(  \Psi_{j}\nabla_{\mathbf{r}}\bar{\Psi}_{j}\right)
^{2}\right]  \right] \nonumber\\
&  \left.  -\frac{\sqrt{m_{1}m_{2}}\gamma}{4\pi}\left(  \bar{\Psi}_{1}\Psi
_{2}+\bar{\Psi}_{2}\Psi_{1}\right)  \right\}  . \label{FGL2}%
\end{align}
Here, the function $\Omega_{s,j}$ formally coincides with the saddle-point
thermodynamic potential for the imbalanced Fermi gas,%
\begin{align}
\Omega_{s,j}  &  =-\int\frac{d\mathbf{k}}{\left(  2\pi\right)  ^{3}}\left[
\frac{1}{\beta}\ln\left(  2\cosh\beta E_{\mathbf{k},j}+2\cosh\beta\zeta
_{j}\right)  \right. \nonumber\\
&  \left.  -\xi_{\mathbf{k},j}-\frac{m_{j}\left\vert \Psi_{j}\right\vert ^{2}%
}{k^{2}}\right]  -\frac{m_{j}\left\vert \Psi_{j}\right\vert ^{2}}{4\pi a_{j}},
\label{Ws}%
\end{align}
where $E_{\mathbf{k},j}=\sqrt{\xi_{\mathbf{k},j}^{2}+\left\vert \Psi
_{j}\right\vert ^{2}}$ is the Bogoliubov excitation energy, $\xi
_{\mathbf{k},j}=\frac{k^{2}}{2m_{j}}-\mu_{j}$ is the free-fermion energy, and
the chemical potentials for the imbalanced fermions are $\mu_{j}=\left(
\mu_{j,\uparrow}+\mu_{j,\downarrow}\right)  /2$ and $\zeta_{j}=\left(
\mu_{j,\uparrow}-\mu_{j,\downarrow}\right)  /2$. However, the order parameter
$\Psi_{j}$ entering this thermodynamic potential is coordinate-dependent. The
parameter $\gamma$ describes the strength of coupling between two bands. The
coefficients $\mathcal{C}_{j},$ $\mathcal{D}_{j}$ and $\mathcal{E}_{j}$
derived in Ref. \cite{ExtGL} are%
\begin{align}
\mathcal{C}_{j}  &  =\int d\mathbf{k}\frac{k^{2}\left[  f_{2}\left(
\beta,E_{\mathbf{k},j},\zeta_{j}\right)  -4\xi_{k}^{2}\left\vert \Psi
_{j}\right\vert ^{2}f_{4}\left(  \beta,E_{\mathbf{k},j},\zeta_{j}\right)
\right]  }{24\pi^{3}m_{j}},\label{c}\\
\mathcal{D}_{j}  &  =\int\frac{d\mathbf{k}}{\left(  2\pi\right)  ^{3}}%
\frac{\xi_{\mathbf{k}}}{\left\vert \Psi_{j}\right\vert ^{2}}\left[
f_{1}\left(  \beta,\xi_{\mathbf{k},j},\zeta_{j}\right)  -f_{1}\left(
\beta,E_{\mathbf{k},j},\zeta_{j}\right)  \right]  ,\label{d}\\
\mathcal{E}_{j}  &  =2\left\vert \Psi_{j}\right\vert ^{2}\int\frac
{d\mathbf{k}}{\left(  2\pi\right)  ^{3}}\frac{k^{2}}{3m_{j}}\xi_{\mathbf{k}%
,j}^{2}~f_{4}\left(  \beta,E_{\mathbf{k},j},\zeta_{j}\right)  . \label{EF}%
\end{align}
The functions $f_{p}\left(  \beta,\varepsilon,\zeta\right)  $ are the
Matsubara sums:
\begin{equation}
f_{p}\left(  \beta,\varepsilon,\zeta\right)  =\sum_{n=-\infty}^{\infty}%
\frac{1}{\left(  \left(  \omega_{n}-i\zeta\right)  ^{2}+\varepsilon
^{2}\right)  ^{p}}. \label{MF2}%
\end{equation}
They can be analytically expressed, e.~g., using the recurrence relations:%
\begin{align}
f_{1}\left(  \beta,\varepsilon,\zeta\right)   &  =\frac{1}{2\varepsilon}%
\frac{\sinh\beta\varepsilon}{\cosh\beta\varepsilon+\cosh\beta\zeta
},\label{msum}\\
f_{p+1}\left(  \beta,\varepsilon,\zeta\right)   &  =-\frac{1}{2p\varepsilon
}\frac{\partial f_{p}\left(  \beta,\varepsilon,\zeta\right)  }{\partial
\varepsilon}.
\end{align}

In order to analytically compare the results of the present approach with the
known GL method near $T_{c}$, we use the results of Ref. \cite{SadeMeloPRL71},
which represents the limiting case of the present approach when $T\rightarrow
T_{c}$ (for a one-band system and without imbalance). For temperatures near
$T_{c}$, the order parameter is small. Thus the coordinate-dependent
thermodynamic potential $\Omega_{s,j}$ is expanded in powers of $\left\vert
\Psi_{j}\right\vert ^{2}$ up to the quartic order, and the coefficients
$\mathcal{C}_{j},$ $\mathcal{D}_{j}$ and $\mathcal{E}_{j}$ are kept for
$\Psi_{j}=0$. In this case, the GL-like free energy (\ref{FGL2}) is reduced to
the TDGL free energy of Ref. \cite{SadeMeloPRL71}, except for the coefficient
$\mathcal{D}$, which appears to be real in the present approach. The reason
{\normalsize for} this difference consists in the following. The imaginary
part in $\mathcal{D}$ appears in Ref. \cite{SadeMeloPRL71} when the gradient
expansion is performed at $\Psi=0$. On the contrary we perform the summations
of the whole series in powers of the order parameter \emph{before} taking the
limit $T\rightarrow T_{c}$, {\normalsize indicating} therefore that
{\normalsize the} appearance of {\normalsize an} imaginary part of
$\mathcal{D}$ depends on a sequence of the limits $\Psi\rightarrow0$ and
$T\rightarrow T_{c}$.

\section{Results}

\label{sec:results}

First, we illustrate a difference of the temperature behavior of the
coefficients of the GL-like free energy (\ref{FGL2}) compared to the
coefficients of the TDGL equation of Ref. \cite{SadeMeloPRL71}.%

%TCIMACRO{\FRAME{fhFU}{3.2235in}{2.4492in}{0pt}{\Qcb{The coefficient $C$
%calculated, as a function of temperature, within the extended TDGL formalism
%(solid curves) and within the TDGL theory of Ref. \cite{SadeMeloPRL71} (dashed
%curves).}}{\Qlb{figs:coefC}}{ktd-vortexviii-2013-fig1.eps}%
%width 3.2235in;  height 2.4492in;  depth 0pt;  original-width 7.1026in;
%original-height 5.382in;  cropleft "0";  croptop "1";  cropright "1";
%filename 'KTD-VortexVIII-2013-Fig1.eps';file-properties "XNPEU";}} }%
%BeginExpansion
\begin{figure}
[h]
\begin{center}
\includegraphics[
height=2.4492in,
width=3.2235in
]%
{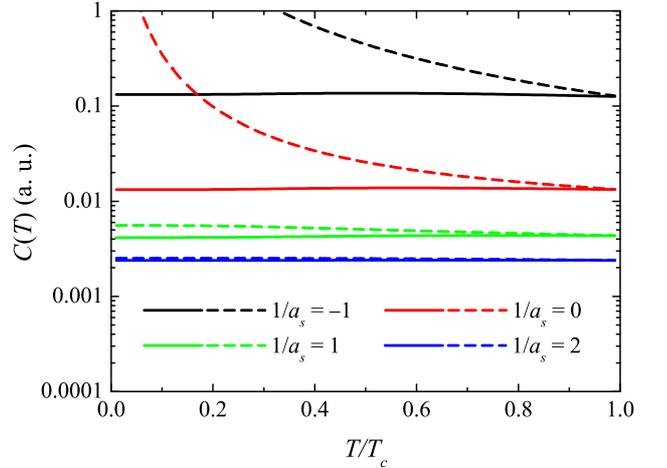}%
\caption{The coefficient $C$ calculated, as a function of temperature, within
the extended TDGL formalism (solid curves) and within the TDGL theory of Ref.
\cite{SadeMeloPRL71} (dashed curves).}%
\label{figs:coefC}%
\end{center}
\end{figure}
%EndExpansion

In Fig. \ref{figs:coefC}, the coefficient $\mathcal{C}$ is plotted as a
function of temperature for several values of the inverse scattering length
$1/a_{s}$ and compared with the coefficient $c$ of Ref. \cite{SadeMeloPRL71}.
Both coefficients analytically tend to the same values when $T\rightarrow
T_{c}$. The temperature behavior of the coefficient $\mathcal{C}$ drastically
differs from that for the corresponding coefficient $c$ within the GL approach
\cite{SadeMeloPRL71}. In the whole range of the BCS-BEC crossover, the
coefficient $\mathcal{C}$ only slightly varies when $T$ goes from $T_{c}$ to
zero. When increasing the inverse scattering length, the range of
temperatures, where $\mathcal{C}$ and $c$ are rather close to each other,
gradually broadens. In the molecular (BEC) regime, both solutions tend to one
and the same limit for all $T\leq T_{c}$. On the contrary, for $1/a_{s}\leq0$,
(i, e., at the BCS side and at unitarity) $c$ rapidly increase when decreasing
temperature (except for the BEC case), and even diverges at $T\rightarrow0$.
These results confirm the fact that the standard GL approach becomes
inapplicable at low temperatures.

Vortices are studied in the present work using the variational method. The
deviations of the order parameters $\Psi_{j}$ from a uniform equilibrium
background can be represented through the product of a uniform background
amplitude $\Delta_{j}\equiv\left\vert \Psi_{j}^{bulk}\right\vert $
{\normalsize with} the amplitude modulation function $f_{j}\left(
\mathbf{r},\tau\right)  $ and the phase factor $e^{i\theta_{j}\left(
\mathbf{r},\tau\right)  }$:
\begin{equation}
\Psi_{j}=\Delta_{j}\cdot f_{j}\left(  \mathbf{r},\tau\right)  e^{i\theta
_{j}\left(  \mathbf{r},\tau\right)  }. \label{Psi}%
\end{equation}
The coefficients $\mathcal{D}_{j},\mathcal{C}_{j},\mathcal{E}_{j}$ are kept
with the bulk values of the order parameter. Thus the time and space
dependence are taken in leading order through the derivatives. This is in line
with the gradient-expansion approximation which was already kept when deriving
(\ref{FGL2}).

Further on, we introduce the notations:%
\begin{align}
\rho_{j}^{\left(  qp\right)  }  &  =\frac{\left(  \mathcal{C}_{j}%
-2\mathcal{E}_{j}\right)  \left\vert \Delta_{j}\right\vert ^{2}}{m_{j}%
},\label{n2}\\
\rho_{j}^{\left(  sf\right)  }  &  =\frac{\left(  \mathcal{C}_{j}%
+2\mathcal{E}_{j}\right)  \left\vert \Delta_{j}\right\vert ^{2}}{m_{j}}.
\label{n4}%
\end{align}
{\normalsize The parameter }$\rho_{j}^{\left(  sf\right)  }$ {\normalsize is
the superfluid density, and }$\rho_{j}^{\left(  qp\right)  }$ {\normalsize is
the quantum pressure coefficient.}

Using (\ref{Psi}), we arrive at the following variational GL-like free energy
functional,%
\begin{align}
F  &  =\frac{1}{\beta}\int_{0}^{\beta}d\tau\int d\mathbf{r}\left\{
\sum_{j=1,2}\left[  \Omega_{s,j}\left(  w_{j}\right)  +i\mathcal{D}%
_{j}\left\vert \Delta_{j}\right\vert ^{2}~f_{j}^{2}\frac{\partial\theta_{j}%
}{\partial\tau}\right.  \right. \nonumber\\
&  \left.  +\frac{1}{2}\rho_{j}^{\left(  qp\right)  }\left(  \nabla
f_{j}\right)  ^{2}+\frac{1}{2}\rho_{j}^{\left(  sf\right)  }f_{j}^{2}\left(
\nabla\theta_{j}\right)  ^{2}\right] \nonumber\\
&  \left.  -\frac{\sqrt{m_{1}m_{2}}\gamma}{2\pi}\Delta_{1}\Delta_{2}%
~f_{1}f_{2}\cos\left(  \theta_{2}-\theta_{1}\right)  \right\}  . \label{FF}%
\end{align}
It describes, in principle, not only the stationary states but also the
time-dependent Josephson physics for a two-band system due to the phase
difference $\theta_{2}\left(  \mathbf{r},\tau\right)  -\theta_{1}\left(
\mathbf{r},\tau\right)  $.%

%TCIMACRO{\FRAME{fhFU}{3.298in}{4.473in}{0pt}{\Qcb{(\QTR{em}{a}) Healing
%lengths for a two-band superfluid fermion system as a function of temperature
%for different values of the coupling parameter $\gamma$; (\QTR{em}{b}) the
%ratio of the healing lengths $\xi_{2}/\xi_{1}$.}}{\Qlb{figs:healing}%
%}{ktd-vortexviii-2013-fig2.eps}{\special{ language "Scientific Word";
%type "GRAPHIC";  maintain-aspect-ratio TRUE;  display "ICON";
%valid_file "F";  width 3.298in;  height 4.473in;  depth 0pt;
%original-width 6.1716in;  original-height 8.387in;  cropleft "0";
%filename 'KTD-VortexVIII-2013-Fig2.eps';file-properties "XNPEU";}} }%
%BeginExpansion
\begin{figure}
[h]
\begin{center}
\includegraphics[
height=4.473in,
width=3.298in
]%
{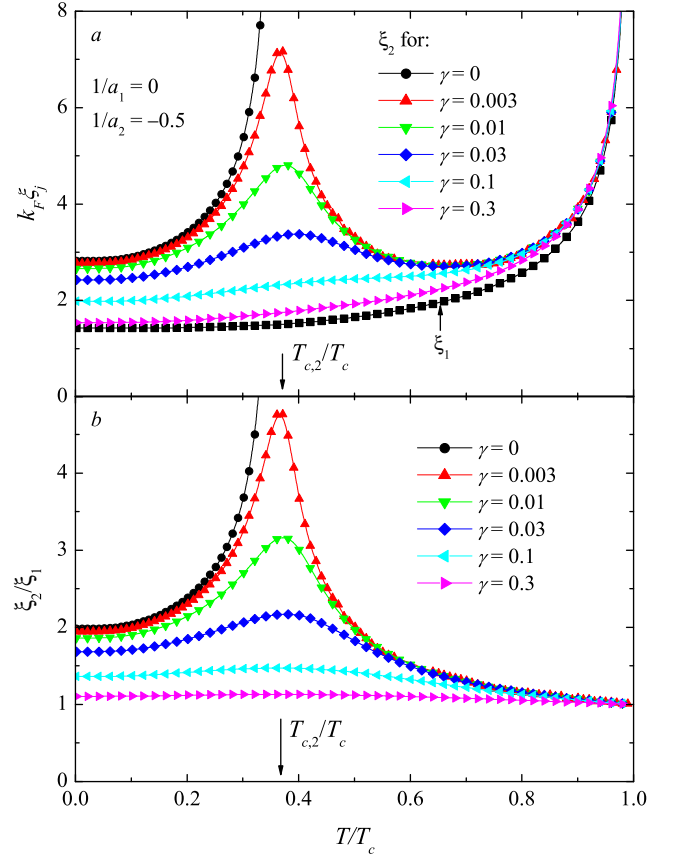}%
\caption{(\emph{a}) Healing lengths for a two-band superfluid fermion system
as a function of temperature for different values of the coupling parameter
$\gamma$; (\emph{b}) the ratio of the healing lengths $\xi_{2}/\xi_{1}$.}%
\label{figs:healing}%
\end{center}
\end{figure}
%EndExpansion

{\normalsize In the present work, the amplitude modulation function for a
vortex was used in the form}%
\begin{equation}
f\left(  r,\xi\right)  =\tanh\left(  \frac{r}{\sqrt{2}\xi}\right)  ,
\label{ampl}%
\end{equation}
{\normalsize with the healing length }$\xi${\normalsize .} The healing lengths
are determined minimizing the free energy (\ref{FF}). In Fig.
\ref{figs:healing}~\emph{a}, the healing lengths for a vortex in a two-band
system are plotted as a function of temperature for the inverse scattering
lengths (in units of the Fermi wave vector $k_{F}$) $1/a_{1}=0$,
$1/a_{2}=-0.5$, and for different values of the coupling parameter $\gamma$.
The healing length for the \textquotedblleft stronger\textquotedblright\ band,
$\xi_{1}$, extremely weakly depends on $\gamma$. The healing length for the
\textquotedblleft weaker\textquotedblright\ band, $\xi_{2}$, demonstrates the
\textquotedblleft hidden criticality\textquotedblright\ discussed in our
manuscript and in Ref. \cite{Komendova}. At zero interband coupling, each of
two subsystems (the \textquotedblleft stronger\textquotedblright\ and
\textquotedblleft weaker\textquotedblright\ bands) is characterized by its own
critical temperature $T_{c,j}$ and healing length $\xi_{j}$, which tends to
infinity at $T\rightarrow T_{c,j}$. When the Josephson interband coupling is
nonzero but sufficiently weak, we can see a fingerprint of the phase
transition for a \textquotedblleft weaker\textquotedblright\ band as a peak of
the healing length $\xi_{2}$ at $T\approx T_{c,2}$.

When comparing the healing lengths for a vortex calculated in the present work
note with those calculated in Ref. \cite{ExtGL} using the model fermion system
near a hard wall, we see a qualitative agreement between the healing lengths
determined by these two methods. However, there is some quantitative
difference between these healing lengths.

In Fig. \ref{figs:healing}~\emph{b}, the ratio $\xi_{2}/\xi_{1}$ is plotted
for the same parameters as in Fig. 1. We can note on a remarkable similarity
between these results and those shown in Fig. 2~(\emph{c}) of Ref.
\cite{Komendova} for a two-band superconductor using BdG equations. The ratio
$\xi_{2}/\xi_{1}$ starts from a value $\xi_{2}/\xi_{1}>1$ at zero temperature,
exhibits a peak near the critical temperature for a \textquotedblleft
weaker\textquotedblright\ band $T_{c,2}$, and tends to 1 when $T\rightarrow
T_{c}$ (which is very close to $T_{c,1}$).

\section{Conclusions}

\label{sec:conclusions}

In summary, we re-formulated the path-integral approach for interacting Fermi
gases \cite{SadeMeloPRL71} to the case of a two-band system. The
Hubbard-Stratonovich transformation and the integration over the fermion
fields lead to an effective bosonic action with Josephson interband coupling.
The gradient expansion of the effective bosonic action results in the GL-like
free energy functional in which the amplitude of the pair field is not a small
parameter. Therefore the obtained free energy represents an extension of the
Ginzburg-Landau formalism to temperatures below $T_{c}$. The range of
applicability of the gradient expansion is determined by the same conditions
as for the standard GL approach, where that expansion is also used. Thus the
present extended GL-like method is valid under the same conditions as the GL
approach -- but in a wider temperature range.

As an example, the method has been tested for vortices in a two-band system of
ultracold fermions. It has been shown that the \textquotedblleft hidden
criticality\textquotedblright\ far below $T_{c}$, treated previously using the
BdG equations \cite{Komendova} is captured by the extended GL-like approach.
Because of the validity of the present approach at temperatures far below
$T_{c}$, it can find a wide spectrum of applications, e. g., for the analysis
of distributions of trapped fermionic atoms, vortices, solitons and other
spatially non-uniform phenomena in ultracold Fermi gases.


\begin{thebibliography}{99}                                                                                               %


\bibitem[*]{A1}This work was presented at the 8$^{\text{th}}$ International
Conference \textquotedblleft Vortex Matter in Nanostructured
Superconductors\textquotedblright,\ September 12 -- 26, 2013, Rhodes, Greece.

\bibitem {Iskin2007}M. Iskin and C. A. R. S\'{a} de Melo, Phys. Rev. A
\textbf{76}, 013601 (2007).

\bibitem {Sam2010}A. V. Samokhvalov, A. S. Mel'nikov, and A. I. Buzdin, Phys.
Rev. B \textbf{82}, 174514 (2010).

\bibitem {Ozawa}T. Ozawa and G. Baym, Phys. Rev. Lett. \textbf{110}, 085304 (2013).

\bibitem {Tieleman}O. Tieleman, O. Dutta, M. Lewenstein, and A. Eckardt, Phys.
Rev. Lett. \textbf{110}, 096405 (2013).

\bibitem {KoganPRB83}V. G. Kogan and J. Schmalian, Phys. Rev. B \textbf{83},
054515 (2011).

\bibitem {Shanenko}A.A. Shanenko, M.V. Milosevic, F.M. Peeters, A.V. Vagov,
Phys. Rev. Lett. \textbf{106}, 047005 (2011).

\bibitem {Shanenko2}A. Vagov, A. A. Shanenko, M. V. Milo\v{s}evi\'{c}, V. M.
Axt, and F. M. Peeters, Phys. Rev. B 86, 144514 (2012).

\bibitem {Orlova}N.V. Orlova, A.A. Shanenko, M.V. Milo\v{s}evi\'{c}, F.M.
Peeters, A.V. Vagov, V.M. Axt, Phys. Rev. B \textbf{87}, 134510 (2013).

\bibitem {Babaev2011}M. Silaev and E. Babaev, Phys. Rev. B \textbf{84}, 094515 (2011).

\bibitem {BabaevPRB86}E. Babaev and M. Silaev, Phys. Rev. B\ \textbf{86},
016501 (2012).

\bibitem {Babaev2012-2}M. Silaev and E. Babaev, Phys. Rev. B \textbf{85},
134514 (2012)

\bibitem {ExtGL}S. N. Klimin, J. Tempere and J. T. Devreese (\emph{to be
published}); arXiv:1309.1421 (2013).

\bibitem {Aranson}I. S. Aranson and L. Kramer, Rev. Mod. Phys. 74, \textbf{99} (2002)

\bibitem {Liao}R. Liao and J. Brand, Phys. Rev. A \textbf{83}, 041604 (2011).

\bibitem {Konotop}V. V. Konotop and L. Pitaevskii, Phys. Rev. Lett.
\textbf{93}, 240403 (2004).

\bibitem {Spunt}A. Spuntarelli, L. D. Carr, P. Pieri, and G. C. Strinati, New
J. Phys. \textbf{13}, 035010 (2011).

\bibitem {Scott}R. G. Scott, F. Dalfovo, L. P. Pitaevskii, and S. Stringari,
Phys. Rev. Lett. \textbf{106}, 185301 (2011).

\bibitem {Yefsah}T. Yefsah, A. T. Sommer, M. J. H. Ku, L. W. Cheuk, W. Ji, W.
S. Bakr and M. W. Zwierlein, Nature \textbf{499}, 426 (2013).

\bibitem {Becker}C. Becker, Nature \textbf{499}, 413 (2013).

\bibitem {Pieri}P. Pieri and G. C. Strinati, Phys. Rev. Lett. \textbf{91},
030401 (2003).

\bibitem {Baksmaty}L. O. Baksmaty, H. Lu, C. J. Bolech and H. Pu, New J. Phys.
\textbf{13}, 055014 (2011).

\bibitem {Ohashi}Y. Ohashi and A. Griffin, Phys. Rev A \textbf{72}, 013601 (2005).

\bibitem {Hu}X.-J. Liu, H. Hu, and P. D. Drummond, Phys. Rev A \textbf{75},
023614 (2007).

\bibitem {SadeMeloPRL71}C. A. R. Sa de Melo, M. Randeria, and J.R.
Engelbrecht, Phys. Rev. Lett. \textbf{71}, 3202 (1993).

\bibitem {Komendova}L. Komendov\'{a}, Y. Chen, A. A. Shanenko, M. V.
Milo\v{s}evic, and F. M. Peeters, Phys. Rev. Lett. \textbf{108}, 207002 (2012).
\end{thebibliography}
\end{document}